# Thermal boundary conductance and phonon transmission in hexagonal boron nitride/graphene heterostructures

David B. Brown[1], Thomas L. Bougher[1], Xiang Zhang[2], Pulickel Ajayan[2], Baratunde A. Cola[1,3], Satish Kumar[1]

[1]G. W. Woodruff School of Mechanical Engineering, Georgia Institute of Technology, Atlanta, GA 30332, USA

[2]Department of Materials Science and NanoEngineering, Rice University, Houston, TX 77005, USA

[3]School of Materials Science and Engineering, Georgia Institute of Technology, Atlanta, GA 30332, USA

**Abstract**

Increased power density in modern microelectronics has led to thermal management challenges which can cause degradation in performance and reliability. In many high-power electronic devices, the power consumption and heat removal are limited by the thermal boundary conductance (TBC) at the interfaces of dissimilar materials. Two-dimensional (2D) materials such as graphene and hexagonal boron nitride (h-BN) have attracted interest as a conductor/insulator pair in next-generation devices because of their unique physical properties; however, the thermal transport at the interfaces must be understood to accurately predict the performance of heterostructures composed of these materials. We use time-domain thermoreflectance (TDTR) to estimate the TBC at the interface of h-BN and graphene to be 35.1 MW/m$^2$-K. We compare the phonon transmission and TBC at the h-BN/graphene interface predicted by two different formulations of the diffuse mismatch model (DMM) for anisotropic materials. The piecewise anisotropic DMM model, which uses two different phonon velocities near the center and at edge of the first Brillouin zone, results in better prediction of phonon transmission rates. The phonon transmission and temperature dependence of TBC confirms the flexural branch in *ab*-plane and *c*-plane longitudinal acoustic branch of graphene and h-BN are the dominant contributor when implementing both the A-DMM and PWA-DMM models. The methodology used here can be employed to heterostructures of other 2D materials.

**Introduction**

The isolation of stable, two-dimensional (2D) crystals [1] began a revolution in condensed matter physics and materials science. Graphene, a 2D material made up of a single layer up to a few layers of $sp^2$ bonded carbon atoms, has attracted considerable interest because of its high intrinsic carrier mobility, mechanical strength, thermal conductivity, and optical transparency [2]. Graphene can be stacked with other 2D materials such as insulating hexagonal boron nitride (h-BN) or transition metal dichalcogenides (TMDs) like molybdenum disulfide ($MoS_2$) or tungsten disulfide ($WS_2$) to build layered, van der Waals heterostructures [3]. These hybrid heterostructures introduce compositional and structural diversities to further enrich the properties and applications of 2D materials [4]. For example, h-BN can be used as a promising substrate for graphene-based field-effect transistors (FETs) and improve mobility of FET's channel [3a, 3b, 3d] compared to $SiO_2$. In addition, graphene/h-BN and graphene/TMD heterostructures showed improved ON-OFF ratio in FET operation due to quantum tunneling [3c, 3f]. The thermoelectric properties of graphene/h-BN heterostructures have also been investigated [5].

To date, the most popular approach to creating 2D material heterostructures has been mechanical stacking of exfoliated or chemical vapor deposition (CVD) grown layers [3a-i, 3k, 6]; however, direct, sequential CVD growth [7] or epitaxial growth on exfoliated 2D layers [8] are also possible. Regardless of preparation method, different stacking arrangements in graphene/h-BN vertical heterostructures are possible resulting in different electronic and phononic properties [9]. Heat dissipation from atomically-thin 2D layers is limited by interfacial transport [10] and makes them an ideal material system for the study of interfacial thermal transport. A fundamental understanding of phonon transport and estimation of the thermal boundary conductance (TBC), also known as Kapitza conductance [11], at the interfaces in 2D material heterostructures is critical to the design process for improving heat dissipation in these devices. Nevertheless, thermal transport across the interfaces in van der Waals heterostructures is still not well understood but is required to keep the device temperature below threshold and enable energy efficient operation. Also, interface quality can vary from sample to sample and across samples based on preparation method making it difficult to obtain an intrinsic measurement. Ultimately, proper control and characterization of the thermal interfaces in layered heterostructures is crucial for practical device applications.

The TBC at graphene/h-BN interfaces have been reported recently [9f, 12]. Using first principle atomistic Green's function (AGF) simulations, Mao et al. [12a] reported a room temperature (RT) TBC of 187 MW/m$^2$-K for a multilayer graphene/multilayer h-BN structure. Zhang et al. [12c] estimated the TBC at graphene nanoribbon/h-BN bilayer structure to be 5 MW/m$^2$-K at RT using classical molecular dynamics (MD) simulations. Yan et al. [9f] used first principles simulations to study the effect of stacking arrangement on TBC for monolayer graphene sandwiched between layers of h-BN. The RT TBC values reported in this study ranged from 30 – 50 MW/m$^2$-K. The first experimental measurement was performed by Chen et al. [12b] using Raman spectroscopy. The reported value of 7.4 MW/m$^2$-K was less than most of the theoretical calculations, which the authors attributed to trapped impurities resulting from the transfer process. Recently, Liu et al. [12d] measured TBC at graphene/h-BN to be 52.2 MW/m$^2$-K using the same Raman technique, while Kim et al. [12e] predicted TBC of 5-10 MW/m$^2$-K in electrically biased graphene FET on h-BN substrate.

Variation in TBC values calculated using different atomistic simulation techniques such as AGF and MD may be expected due to different assumptions and limitations. However, there is discrepancy in results even when the same measurement technique, Raman spectroscopy, is used. Also, the Raman technique requires a patterning step to form leads to electrically heating the graphene to create a temperature difference. In this work, time-domain thermoreflectance (TDTR), which requires only the deposition of a thin metal film, is used *for the first time* to estimate the TBC at the interfaces of graphene and h-BN. Our measured value of 35.1 MW/m$^2$-K lies in between the previously reported experimental values and also in the range of TBC predicted by first principles density functional theory and AGF based calculations [9f]. We also present the phonon transmission and TBC predicted by two formulations of the diffuse mismatch model (DMM) for anisotropic materials like graphene and h-BN.

**Experiments and Modeling**

Single layer graphene (SLG), with some bilayer islands, and few-layer h-BN grown on Cu foil using separate chemical vapor deposition (CVD) processes [7a] were transferred to the surface of 300 nm thermally-grown SiO$_2$ (measured using a Nanometrics Nanospec 3000 reflectometer) using a poly(methyl methacrylate) (PMMA) polymer support. Prior to transfer to SiO$_2$, the underlying Cu foil was etched in FeCl$_3$, and, following transfer, the PMMA was dissolved in acetone and isopropyl alcohol. Finally, the samples were annealed at 300°C in vacuum (5-10 mTorr) to remove residual PMMA [13] and improve conformity to the substrate [14]. Raman spectroscopy data was acquired using a Renishaw InVia Raman microscope with 180° backscattering geometry and 488 nm Ar$^+$ laser focused using a 50x objective lens (NA=0.5). X-ray photoelectron spectroscopy (XPS) was performed using a Thermo Scientific K-Alpha[+] spectrometer with an Al Kα monochromatic X-ray source (1486.6 eV). In preparation for thermal measurements, the samples were simultaneously coated with Au (3 nm Ti adhesion layer) using electron-beam evaporation to serve as a thermal transducer. Schmidt et al. [15] showed that inclusion of 5 nm Ti adhesion layer nearly doubled the TBC at Al-graphite interface. Therefore, the interfaces considered here are Ti/G/SiO$_2$ or Ti/h-BN/SiO$_2$ despite Au layer as the thermal transducer. The sample geometries used in this study are shown in Figure 1. The actual film thickness of 77 nm (Figure 1) was measured on co-deposited glass slide using a Veeco Dimension 3100 Atomic Force Microscope in tapping mode.

TDTR has become a widely used technique to measure the thermal conductivity of thin films and substrates as well as TBC [16]. Briefly, TDTR is pump-probe optical technique which uses a modulated laser beam (pump) to heat the surface of a sample and an unmodulated beam (probe) to measure the change in optical reflectivity of the surface. Modulation of the pump beam allows the signal to be measured using lock-in amplification. The experimental data is fit to a thermal model [17] in order to extract the thermal properties of interest. In the two-color TDTR setup used in this study, described previously [18], the output of a Spectra Physics Ti:Sapphire ($\lambda$=800 nm, 40 nJ/pulse) laser with ~150 fs pulse width and a repetition rate of ~80 MHz is split into two beam paths (pump and probe) where the pump beam is modulated at a frequency of 8.8 MHz then frequency doubled using a BiBO crystal. The pump and probe strike the surface concentrically at a normal angle of incidence and are focused to $1/e^2$ radii of ~5 and ~3 μm, respectively, using pump and probe powers of 10 and 4 mW, respectively. The arrival time of the probe is delayed up to 5 ns relative to the pump by adjusting its optical length using a double-pass mechanical delay stage to map the decay of the thermoreflectance signal. Monte Carlo (MC) simulations were used to determine uncertainties associated with TBC estimation.

The original derivation of the DMM presented by Swartz and Pohl [11b] assumed diffuse, elastic scattering of phonons at the interface of two materials. A later study by Stoner and Maris [19] showed that the elastic assumption under predicts the TBC compared to experimental measurements. More recent work has taken into account phonon dispersion [20], interfacial mixing [21], surface roughness [22], and inelastic scattering [23] with varying amounts of success. Nevertheless, the DMM remains a useful tool for capturing trends in the phonon transmission across interfaces and because of its simple implementation. For small $\Delta T$, the TBC approaches $TBC = \tau \frac{\partial (\alpha_{12} H_1)}{\partial T}$ where, under the diffuse assumption [11b], $\alpha_{12} = H_2/(H_1 + H_2)$, is the transmission coefficient from material 1 to material 2 and the incident phonon irradiation, $H$ (W/m$^2$), an analog to photon irradiation [24], from a 3D isotropic solid is given by the relationship,

$$H = \frac{1}{4}\sum_j \int_0^{\omega_{\max,j}} \hbar\omega v_{j,1} f(\omega,T) D(\omega) d\omega, \tag{1}$$

where $\omega$ is angular frequency, $v_j$ is the phonon group velocity, $f$ is the Bose-Einstein distribution function ($f(\omega,T) = 1/[\exp(\hbar\omega/k_B T) - 1]$), and $\hbar$ and $k_B$ are the reduced Planck constant and Boltzmann constant, respectively. The summation is carried out over each phonon polarization, and the integration limits, $\omega_{\max,j}$, correspond to the maximum frequency considered in each material. $D(\omega)$ is the phonon density of states (DOS), and under the Debye assumption [25], $\omega = vk$, where $k$ is the wavevector, it is given by $D(\omega) = \omega^2/(2\pi^2 v_j^3)$. Plugging this expression for $D(\omega)$ into Equation 1 it can be shown that,

$$H = \frac{k_B^4}{8\pi^2 \hbar^3}\sum_j \frac{1}{v_j^2} \int_0^{x_{\max,j}} \frac{T^4 x^3}{\exp(x)-1} dx. \tag{2}$$

In equation 2, $x = \hbar\omega/k_B T$. The prefactor $\tau = \left[1 - \frac{1}{2}(\alpha_{12} + \alpha_{21})\right]^{-1}$ is necessary when working in terms of local equivalent equilibrium temperature [26].

An isotropic Debye dispersion (i.e., $\omega = vk$) is generally assumed and may be valid for Ti, but this assumption is not acceptable in the highly anisotropic, 2D graphene and h-BN. Duda et al. [27] accounted for this anisotropy and calculated TBC at a metal/graphite interface by using an effective 2D Debye density of states, $D_{2D,eff}(\omega) = \omega/(2\pi v^2 d)$, where $d$ is the interlayer spacing for graphite. More recently, Chen et al. [28] showed this 2D DMM model greatly overpredicts the TBC and presented a new DMM model using an anisotropic Debye dispersion ($\omega^2 = v_{ab}^2 k_{ab}^2 + v_c^2 k_c^2$, where $k_{ab}^2 = k_a^2 + k_b^2$) referred to here as anisotropic-DMM (A-DMM). The resulting first Brillouin zone (FBZ) is ellipsoidal, as opposed to spherical in the case of an isotropic dispersion, where the major and minor axes correspond to the graphite $ab$- (i.e. basal) and $c$-axis. The authors presented a detailed derivation and analysis where the real quasi-TA and quasi-LA branches of the phonon dispersion [29] are recomposed into two ellipsoids: TL1 and TL2 branches. Along with an additional TA branch, the phonon velocities ($v_{ab,j}$ and $v_{c,j}$) for each branch can be determined from the real phonon dispersion. The phonon irradiation given by Chen et al. [28] is,

$$H_{\text{A-DMM}} = \frac{k_B^4}{8\pi^2\hbar^3}\sum_j \frac{1}{v_{ab,j}^2}\left\{\int_0^{x_{\max,c,j}} \frac{T^4 x^3}{e^x-1}dx + \int_{x_{\max,c,j}}^{x_{\max,ab,j}}\left[\frac{\theta_{D,ab,j}^2\theta_{D,c,j}^2}{\theta_{D,ab,j}^2-\theta_{D,c,j}^2}\frac{T^2 x}{e^x-1} - \frac{\theta_{D,c,j}^2}{\theta_{D,ab,j}^2-\theta_{D,c,j}^2}\frac{T^4 x^3}{e^x-1}\right]dx\right\}, \quad (3)$$

where $\theta_{D,j}$ is the Debye temperature ($\hbar\omega_{D,j}/k_B$). We use the $\omega$ values from Table 1 to estimate $\theta_{D,j}$. The first term in Equation (3) is identical to Equation (1) from the original DMM. The metal/graphite TBC results by Chen et al. [28] showed the model still overpredicts the TBC at metal/graphite interfaces when compared to experiments [15].

An update to the A-DMM reported by Li et al. [30] attempts to resolve any discrepancy caused by input parameters. While Chen et al. [28] used the "secant" method (i.e., the slope of secant line connecting the $\Gamma$ point to the end of the FBZ) to estimate phonon velocity of each branch, this method greatly overpredicts the phonon velocity of the flexural (ZA) branch found in graphite and other layered materials like h-BN. Li et al. [30] instead utilizes the elastic constants to predict the phonon velocities and, cleverly employs a piecewise (PW) linear approximation for the ZA branch specifically. The PW linear approximation is a more accurate representation of the ZA branch because at small wavevector, the ZA branch varies as $k^2$ which differs from TA and LA branches which vary as $k$ [31]. We refer to this model as PW anisotropic DMM (PWA-DMM). In addition, the cutoff frequencies are determined from the real phonon dispersion as opposed to the Debye approximation ($\omega_{D,j} = v_j k_D$). The $c$-axis phonon irradiation ($H_{\text{PWA-DMM}}$) for the PWA-DMM is identical to the A-DMM model for the TA and TL1 branches. Following some derivation, the expression for TL2 branch can be given by,

$$H_{\text{PWA-DMM,TL2}} = \frac{k_B^4}{8\pi^2\hbar^3}\left\{\frac{1}{v_{ab,1}^2}\int_0^{x_{\max,ab,1}}\frac{T^4 x^3}{e^x-1}dx + \frac{1}{v_{ab,2}^2}\left[\int_{x_{\max,ab,1}}^{x_{\max,c}}\frac{T^4 x(x-\Delta x)^2}{e^x-1}dx + \int_{x_{\max,c}}^{x_{\max,ab,2}}\frac{\hbar^2 v_{ab,2}^2 k_{ab,2}^2/k_B^2 T^2 - (x-\Delta x)^2}{\hbar^2 v_{ab,2}^2 k_{ab,2}^2 x^2/k_B^2 T^2 x_{\max,c}^2 - (x-\Delta x)^2}\frac{T^4 x(x-\Delta x)^2}{e^x-1}dx\right]\right\} \quad (4)$$

In Equation 4, $v_{ab,1}$ and $v_{ab,2}$ are the phonon velocities corresponding to the first and second segment of ZA branch using the piecewise linear approximation, respectively, and $\Delta x =$

$\hbar\Delta\omega/k_\text{B}T$ where $\Delta\omega = k_{ab,1}(v_{ab,1} - v_{ab,2})$. $k_{ab,1}$ and $k_{ab,2}$ are wavevectors corresponding to intersection of the two piecewise segments and the cutoff wavevector in the $ab$-plane. The cutoff wave vectors are determined using the relationship $k_{ab}^2 k_c = 6\pi^2 N/V$ and the anisotropy ratio of the real lattice to ensure the correct number of acoustic modes [28]. We implement both models here to calculate the TBC at Ti/G, Ti/ h-BN, and h-BN/G interfaces.

**Results and Discussion**

The optical microscope image in Figure 1d shows a 1x1 mm$^2$ area of SiO$_2$ coated with mostly SLG and h-BN. Figure 3 shows the Raman spectra from the sample used in this study. The graphene sample (Fig. 2a) with G peak at 1592 cm$^{-1}$ (E$_{2g}$ mode near the Γ point) and 2D peak 2703 cm$^{-1}$ (A$_{1g}$ mode near the K point) [32] and intensity ratio I(2D)/I(G) ≈ 2.2 [33] shows our sample is single-layer; however, the shift in peak positions and reduced I(2D)/I(G) ratio suggests some p-type doping [34] previously attributed to residual PMMA [35]. The D peak at 1356 cm$^{-1}$ arises from disorder in the graphene layer. Figure 2b shows the peak in h-BN Raman spectrum blue-shifted to ~1370 cm$^{-1}$, corresponding to the in-plane E$_{2g}$ mode, compared to the characteristic peak at ~1366 cm$^{-1}$ for bulk h-BN [36]. This shift could be caused by stress in the film resulting from the growth process, substrate/interlayer interaction, or crystallite size [37]. A comparison of the graphene and h-BN/graphene Raman spectra (not shown) did not display new peaks in the range of 1200-3200 cm$^{-1}$ which would suggest coupling between the 2D layers [38]. There was simply peak broadening around 1360 cm$^{-1}$ as a result of the h-BN layer. A more extensive Raman study may reveal shear or layer-breathing modes at lower frequencies. The high resolution XPS spectra in Figure 2c and 2d, respectively, show B and N peaks at binding energies of 191 and 398 eV, respectively. From the XPS data, the stoichiometry of our h-BN sample was 1.17:1 (B:N) [39].

The TBC at Ti/G/SiO$_2$ and Ti/h-BN/SiO$_2$ interfaces were 30.6 (+6.2/-4.2) and 33.7 (+9.8/-5.3) MW/m$^2$-K, respectively. The TBC at Ti/h-BN/G/SiO$_2$ interface was 18.3 (+4.0/-2.7) MW/m$^2$-K. The TDTR signals for these samples are compared in Figure 3a. The total interfacial thermal conductance per unit area can be ascribed to the metal/h-BN/G/SiO$_2$ interfaces acting in series, as in the case of a thin film sample between two solids [40]. We, therefore, use a one-dimensional thermal resistance network to estimate the TBC at h-BN/G interface. This method was used previously [16f, 41] where the heat transport across metal/G/SiO$_2$ and metal/G/metal interfaces were treated as the resistances of the decoupled metal/G and G/SiO$_2$ (or G/metal) interfaces acting in series. Zheng et al. [41d] reevaluated this analysis recently suggesting long wavelength phonons may traverse both interfaces through a process similar to the heat transport in superlattices [42].

Nevertheless, we apply the method here in the following manner. Using the relationship, $1/TBC_{\text{Ti/G/SiO2}} = 1/TBC_{\text{Ti/G}} + 1/TBC_{\text{G/SiO2}}$, we can determine $TBC_{\text{Ti/G}}$. Similarly, the $TBC_{\text{G/SiO2}}$ term in this equation can be replaced by $TBC_{\text{h-BN/SiO2}}$ to determine $TBC_{\text{Ti/h-BN}}$. The thermal conductance of the h-BN and SLG layers were much greater than the interfacial TBC and were therefore neglected.

The $TBC_{\text{G/SiO2}}$ and $TBC_{\text{h-BN/SiO2}}$ values were previously reported for SLG (~80 MW/m²-K) [43] and monolayer h-BN (~63 MW/m²-K) [44] using the 3$\omega$ technique. Using these values, the resulting $TBC_{\text{Ti/G}}$ and $TBC_{\text{Ti/h-BN}}$ are 49.6 (+18.7/-10.2) and 73.3 (+69.1/-21.1) MW/m²-K. The uncertainty bounds were determined using the upper/lower limits from the MC simulations for Ti/G/SiO₂ and Ti/h-BN/SiO₂ interfaces. We use these values and formulate a new relationship, $1/TBC_{\text{Ti/h-BN/G/SiO2}} = 1/TBC_{\text{Ti/h-BN}} + 1/TBC_{\text{h-BN/G}} + 1/TBC_{\text{G/SiO2}}$, and estimate $TBC_{\text{h-BN/G}}$ to be 35.1 (+4.5/-4.2) MW/m²-K. The TBC values are summarized in Figure 3b. When compared with previous values in literature, our TBC value is greater than 7.4 MW/m²-K reported by Chen et al. [12b] and 5-10 MW/m²-K reported by Kim et al. [12e] However, our value is less than 52.2 MW/m²-K reported by Liu et al. [12d], which we attribute to surface roughness resulting from the CVD growth process and PMMA residue following the transfer process. Our value is also in similar range as TBC (30 – 50 MW/m²-K) for different lattice stacking configurations predicted using first-principles AGF simulations [9f].

Phonon velocities were calculated using the elastic constants for Ti [29], graphite [45], and h-BN [46]. The cutoff frequencies for each branch were determined from the published dispersion relations [31, 47], and the Debye temperature for each branch corresponds to these frequencies. We follow Chen et al. [28] and determine the cutoff wavevectors using the relationship $k_{ab}^2 k_c = 6\pi^2 N/V$ and the anisotropy ratio of the real lattice ensuring the correct number of acoustic modes. We also unfold the dispersion relation along $c$-axis because of the relatively high velocity of optical modes in that direction. Input parameters for both models are listed in Table 1, where $v_{ab,2}$, $\omega_{ab,1}$, and $k_{ab,1}$ are not used in the A-DMM model. Unlike Li et al. [30], we use the same input parameters for both models for direct comparison.

The DMM does not consider the quality of the interface (e.g., bonding, roughness), which varies from sample to sample; therefore, we hold $\alpha_{12}$ constant and determined its value for each interface by fitting both DMM models to our RT TDTR measurements (Table 2). As a result, only the phonon irradiation from material 1 (e.g., Ti in the case of Ti/G and Ti/h-BN interface) needs to be considered [11b]. Thus, when utilizing the fitted values, $\alpha_{12,\text{fit}}$, the A-DMM and PWA-DMM models differ from each other, and from the original DMM (Equation 2), only when considering the h-BN/G interface. The $\alpha_{12,\text{fit}}$ values listed in Table 2 are very insightful. As expected, $\alpha_{12,\text{fit}}$ for the Ti/G and Ti/h-BN are identical for the A-DMM and PWA-DMM models. They are also similar order of magnitude (~$10^{-2}$) for $\alpha_{12,\text{fit}}$ at metal/graphite interfaces [15, 41a] reported in previous studies. We must point out that Schmidt et al. [15] assumed a sine-type (or Born-von Karman) [24] dispersion for metals and effective 2D Debye density of states [27] in graphite. Also, the velocities of each phonon polarization were lumped into a single, average velocity. Koh et al. [41a] used a linear (Debye) dispersion for Au. $\alpha_{12,\text{fit}}$ for h-BN/G interface predicted by the PWA-DMM was nearly an order of magnitude larger than the value predicted by the A-DMM (see Table 2). The reason for this discrepancy is discussed below.

The ratio of $\alpha_{12}$, calculated using the phonon irradiation (Equations 3 and 4) and the relationship $\alpha_{12} = H_2/(H_1 + H_2)$, to $\alpha_{12,\text{fit}}$ values in Table 2 is compared in Figure 4a. While Li et al. [30] makes an elastic assumption in determining $\alpha_{12}(\omega)$, we assume inelastic scattering [48] in accordance with the A-DMM model when computing $\alpha_{12}(T)$ allowing phonons of all frequencies in h-BN and graphene to participate. For the h-BN/G interface, $\alpha_{12}$ is expected to be close to 0.5 for both models due to the similar vibrational properties of graphene and h-BN [31, 47b]. The graph of $\alpha_{12}/\alpha_{12,\text{fit}}$ in Figure 4a shows a weak temperature dependence above 200 K for both the A-DMM model (solid lines) and PWA-DMM (dashed lines). The discrepancy between $\alpha_{12}$ and $\alpha_{12,\text{fit}}$ was much larger for A-DMM compared to PWA-DMM model. At RT $\alpha_{12}/\alpha_{12,\text{fit}}$ for Ti/G and Ti/h-BN interfaces are 14.9 and 7.27, respectively, using the A-DMM model, while there is much better agreement for the PWA-DMM (1.36 and 1.55 for Ti/G and Ti-h-BN, respectively).

The high $\alpha_{12}/\alpha_{12,\text{fit}}$ ratio for the A-DMM for Ti/G and Ti/h-BN interface arises from the much higher phonon irradiation in graphene and h-BN compared to PWA-DMM. The phonon irradiation is proportional to $v_{ab}^{-2}$, thus the assumption of a constant $v_{ab}$ for TL2 branch, which contributes most to the irradiation [28, 30], results in much higher calculated $\alpha_{12}$ value for A-DMM model. This is the phonon focusing [49] effect whereby cross-plane TBC can be increased with a reduction in in-plane phonon velocity. The same is true for h-BN/G interface where $\alpha_{12}/\alpha_{12,\text{fit}}$ was 30.3 and 1.57 for the A-DMM and PWA-DMM models, respectively. The transmission coefficient for each phonon branch, $\alpha_{12,j}$, is shown in Figure 4b and further reinforces the importance of the TL2 branch.

Finally, the TBC predicted using $\alpha_{12,\text{fit}}$ are shown in Figure 5 along with TBC for Ti/G, Ti/h-BN, and h-BN/G from our TDTR results. Various literature results for h-BN/G [12b, 12d, 12e], metal/G [16f, 41d, 50] and metal/graphite [15] interfaces are also shown for comparison. The discrepancy between h-BN/G results for the A-DMM and PWA-DMM models at low temperatures arises from the assumption of constant $\alpha_{12}$. The phonon characteristic wavelength varies as $T^{-1}$ resulting in higher $\alpha_{12}$ at low temperatures where phonon characteristic wavelength is much larger than surface roughness leading to decreased scattering [11b]. This behavior is captured by the PWA-DMM model but not the A-DMM. Interestingly, the TBC at h-BN/G interface is constant above 200 K for both models. TBC is expected to increase below the Debye temperature [11b, 51], which is greater than 1000 K [52] for both h-BN and graphene. The observed trend with temperature may be a result of the ZA branch in $ab$-axis and LA branch in $c$-axis (i.e., TL2 branch) being the dominant contributor to TBC for both A-DMM and PWA-DMM models. The maximum frequency of vibrations (Table 1) for the two branches correspond to Debye temperatures of 764 and 174 K, respectively, but the contribution from both remain constant above 200 K.

**Conclusion**

We have estimated the TBC at h-BN/G interface using a series thermal resistor network coupled with TDTR measurements at Ti/G/SiO$_2$, Ti/h-BN/SiO$_2$, and Ti/h-BN/G/SiO$_2$ interfaces. However, since h-BN and graphene have similar physical structure and acoustic properties the h-BN/G TBC may be increased by improving sample quality. We compare the phonon transmission using two forms of the DMM for anisotropic materials. The A-DMM model predicts a higher phonon irradiation thus higher transmission coefficient due to the assumption of constant velocity of ZA mode across entire the FBZ. The PWA-DMM model uses two different phonon velocities near the center and at edge of the FBZ resulting in better prediction of phonon transmission. The phonon transmission and temperature dependence of TBC confirms the ZA branch along the $ab$-axis and the LA branch along the $c$-axis of graphene and h-BN are the dominant contributor when implementing both the A-DMM and PWA-DMM model. This methodology can be extended to other 2D heterostructures to analyze the TBC at the interfaces of 2D layers.

**Acknowledgements**

D.B.B. was supported by a National Science Foundation Graduate Research Fellowship under Grant No. DGE-1650044. Any opinion, findings, and conclusions or recommendations expressed

**References**

1. Novoselov, K.S., et al., *Two-dimensional atomic crystals.* Proceedings of the National Academy of Sciences of the United States of America, 2005. **102**(30): p. 10451-10453.
2. (a) Novoselov, K.S., et al., *Electric field effect in atomically thin carbon films.* Science, 2004. **306**(5696): p. 666-669; (b) Balandin, A.A., et al., *Superior thermal conductivity of single-layer graphene.* Nano Letters, 2008. **8**(3): p. 902-907; (c) Bolotin, K.I., et al., *Ultrahigh electron mobility in suspended graphene.* Solid State Communications, 2008. **146**(9-10): p. 351-355; (d) Lee, C., et al., *Measurement of the elastic properties and intrinsic strength of monolayer graphene.* Science, 2008. **321**(5887): p. 385-388; (e) Nair, R.R., et al., *Fine structure constant defines visual transparency of graphene.* Science, 2008. **320**(5881): p. 1308-1308.
3. (a) Dean, C.R., et al., *Boron nitride substrates for high-quality graphene electronics.* Nature Nanotechnology, 2010. **5**(10): p. 722-726; (b) Mayorov, A.S., et al., *Micrometer-Scale Ballistic Transport in Encapsulated Graphene at Room Temperature.* Nano Letters, 2011. **11**(6): p. 2396-2399; (c) Britnell, L., et al., *Field-Effect Tunneling Transistor Based on Vertical Graphene Heterostructures.* Science, 2012. **335**(6071): p. 947-950; (d) Lee, K.H., et al., *Large-Scale Synthesis of High-Quality Hexagonal Boron Nitride Nanosheets for Large-Area Graphene Electronics.* Nano Letters, 2012. **12**(2): p. 714-718; (e) Britnell, L., et al., *Strong Light-Matter Interactions in Heterostructures of Atomically Thin Films.* Science, 2013. **340**(6138): p. 1311-1314; (f) Georgiou, T., et al., *Vertical field-effect transistor based on graphene-WS2 heterostructures for flexible and transparent electronics.* Nature Nanotechnology, 2013. **8**(2): p. 100-103; (g) Yu, W.J., et al., *Vertically stacked multi-heterostructures of layered materials for logic transistors and complementary inverters.* Nature Materials, 2013. **12**(3): p. 246-252; (h) Roy, K., et al., *Graphene-MoS2 hybrid structures for multifunctional photoresponsive memory devices.* Nature Nanotechnology, 2013. **8**(11): p. 826-830; (i) Kretinin, A.V., et al., *Electronic Properties of Graphene Encapsulated with Different Two-Dimensional Atomic Crystals.* Nano Letters, 2014. **14**(6): p. 3270-3276; (j) Roy, T., et al., *Field-Effect Transistors Built from All Two-Dimensional Material Components.* Acs Nano, 2014. **8**(6): p. 6259-6264; (k) Withers, F., et al., *Light-emitting diodes by band-structure engineering in van der Waals heterostructures.* Nature Materials, 2015. **14**(3): p. 301-306.
4. (a) Geim, A.K. and I.V. Grigorieva, *Van der Waals heterostructures.* Nature, 2013. **499**(7459): p. 419-425; (b) Novoselov, K.S., et al., *2D materials and van der Waals heterostructures.* Science, 2016. **353**(6298); (c) Liu, Y., et al., *Van der Waals heterostructures and devices.* Nature Reviews Materials, 2016. **1**(9).
5. (a) Chen, C.C., et al., *Thermoelectric transport across graphene/hexagonal boron nitride/graphene heterostructures.* Nano Research, 2015. **8**(2): p. 666-672; (b) D'Souza, R. and S. Mukherjee, *Thermoelectric transport in graphene/h-BN/graphene heterostructures: A computational study.* Physica E-Low-Dimensional Systems & Nanostructures, 2016. **81**: p. 96-101.
6. Mudd, G.W., et al., *High Broad-Band Photoresponsivity of Mechanically Formed InSe-Graphene van der Waals Heterostructures.* Advanced Materials, 2015. **27**(25): p. 3760-3766.
7. (a) Liu, Z., et al., *Direct Growth of Graphene/Hexagonal Boron Nitride Stacked Layers.* Nano Letters, 2011. **11**(5): p. 2032-2037; (b) Shi, Y.M., et al., *van der Waals Epitaxy of*

*MoS2 Layers Using Graphene As Growth Templates.* Nano Letters, 2012. **12**(6): p. 2784-2791; (c) Wang, M., et al., *A Platform for Large-Scale Graphene Electronics - CVD Growth of Single-Layer Graphene on CVD-Grown Hexagonal Boron Nitride.* Advanced Materials, 2013. **25**(19): p. 2746-2752; (d) Zhang, C.H., et al., *Direct growth of large-area graphene and boron nitride heterostructures by a co-segregation method.* Nature Communications, 2015. **6**.

8. (a) Yang, W., et al., *Epitaxial growth of single-domain graphene on hexagonal boron nitride.* Nature Materials, 2013. **12**(9): p. 792-797; (b) Lin, Y.C., et al., *Atomically Thin Heterostructures Based on Single-Layer Tungsten Diselenide and Graphene.* Nano Letters, 2014. **14**(12): p. 6936-6941; (c) Yan, A.M., et al., *Direct Growth of Single- and Few-Layer MoS2 on h-BN with Preferred Relative Rotation Angles.* Nano Letters, 2015. **15**(10): p. 6324-6331.

9. (a) Giovannetti, G., et al., *Substrate-induced band gap in graphene on hexagonal boron nitride: Ab initio density functional calculations.* Physical Review B, 2007. **76**(7); (b) Decker, R., et al., *Local Electronic Properties of Graphene on a BN Substrate via Scanning Tunneling Microscopy.* Nano Letters, 2011. **11**(6): p. 2291-2295; (c) Sachs, B., et al., *Adhesion and electronic structure of graphene on hexagonal boron nitride substrates.* Physical Review B, 2011. **84**(19); (d) Xue, J.M., et al., *Scanning tunnelling microscopy and spectroscopy of ultra-flat graphene on hexagonal boron nitride.* Nature Materials, 2011. **10**(4): p. 282-285; (e) Slotman, G.J., et al., *Phonons and electron-phonon coupling in graphene-h-BN heterostructures.* Annalen Der Physik, 2014. **526**(9-10): p. 381-386; (f) Yan, Z.Q., et al., *Phonon transport at the interfaces of vertically stacked graphene and hexagonal boron nitride heterostructures.* Nanoscale, 2016. **8**(7): p. 4037-4046.

10. Pop, E., *Energy Dissipation and Transport in Nanoscale Devices.* Nano Research, 2010. **3**(3): p. 147-169.

11. (a) Kapitza, P.L., *Heat transfer and superfluidity of helium II.* Journal of Physics-Ussr, 1941. **5**(1-6): p. 59-69; (b) Swartz, E.T. and R.O. Pohl, *Thermal-Boundary Resistance.* Reviews of Modern Physics, 1989. **61**(3): p. 605-668.

12. (a) Mao, R., et al., *Phonon engineering in nanostructures: Controlling interfacial thermal resistance in multilayer-graphene/dielectric heterojunctions.* Applied Physics Letters, 2012. **101**(11); (b) Chen, C.C., et al., *Thermal interface conductance across a graphene/hexagonal boron nitride heterojunction.* Applied Physics Letters, 2014. **104**(8); (c) Zhang, J.C., Y. Hong, and Y.A. Yue, *Thermal transport across graphene and single layer hexagonal boron nitride.* Journal of Applied Physics, 2015. **117**(13); (d) Liu, Y., et al., *Thermal Conductance of the 2D MoS2/h-BN and graphene/h-BN Interfaces.* Scientific Reports, 2017. **7**; (e) Daehee, K., et al., *Energy dissipation mechanism revealed by spatially resolved Raman thermometry of graphene/hexagonal boron nitride heterostructure devices.* 2D Materials, 2018. **5**(2): p. 025009.

13. Lin, Y.C., et al., *Graphene Annealing: How Clean Can It Be?* Nano Letters, 2012. **12**(1): p. 414-419.

14. Huang, B. and Y.K. Koh, *Improved topological conformity enhances heat conduction across metal contacts on transferred graphene.* Carbon, 2016. **105**: p. 268-274.

15. Schmidt, A.J., et al., *Thermal conductance and phonon transmissivity of metal-graphite interfaces.* Journal of Applied Physics, 2010. **107**(10).

16. (a) Cahill, D.G., et al., *Nanoscale thermal transport.* Journal of Applied Physics, 2003. **93**(2): p. 793-818; (b) Cahill, D.G., *Analysis of heat flow in layered structures for time-domain thermoreflectance.* Review of Scientific Instruments, 2004. **75**(12): p. 5119-5122; (c) Koh, Y.K. and D.G. Cahill, *Frequency dependence of the thermal conductivity of* *semiconductor alloys.* Physical Review B, 2007. **76**(7); (d) Schmidt, A.J., X.Y. Chen, and G. Chen, *Pulse accumulation, radial heat conduction, and anisotropic thermal conductivity in pump-probe transient thermoreflectance.* Review of Scientific Instruments, 2008. **79**(11); (e) Hopkins, P.E., et al., *Reduction in the Thermal Conductivity of Single Crystalline Silicon by Phononic Crystal Patterning.* Nano Letters, 2011. **11**(1): p. 107-112; (f) Hopkins, P.E., et al., *Manipulating Thermal Conductance at Metal-Graphene Contacts via Chemical Functionalization.* Nano Letters, 2012. **12**(2): p. 590-595.

17. Liu, J., et al., *Simultaneous measurement of thermal conductivity and heat capacity of bulk and thin film materials using frequency-dependent transient thermoreflectance method.* Review of Scientific Instruments, 2013. **84**(3).
18. (a) Bougher, T.L., et al., *Thermal Boundary Resistance in GaN Films Measured by Time Domain Thermoreflectance with Robust Monte Carlo Uncertainty Estimation.* Nanoscale and Microscale Thermophysical Engineering, 2016. **20**(1): p. 22-32; (b) Brown, D.B., et al., *Oxidation limited thermal boundary conductance at metal-graphene interface.* Carbon, 2018. **139**: p. 913-921.
19. Stoner, R.J. and H.J. Maris, *Kapitza Conductance and Heat-Flow between Solids at Temperatures from 50 to 300 K.* Physical Review B, 1993. **48**(22): p. 16373-16387.
20. Reddy, P., K. Castelino, and A. Majumdar, *Diffuse mismatch model of thermal boundary conductance using exact phonon dispersion.* Applied Physics Letters, 2005. **87**(21).
21. Beechem, T., et al., *Role of interface disorder on thermal boundary conductance using a virtual crystal approach.* Applied Physics Letters, 2007. **90**(5).
22. Hopkins, P.E., et al., *Effects of surface roughness and oxide layer on the thermal boundary conductance at aluminum/silicon interfaces.* Physical Review B, 2010. **82**(8).
23. Hopkins, P.E., J.C. Duda, and P.M. Norris, *Anharmonic Phonon Interactions at Interfaces and Contributions to Thermal Boundary Conductance.* Journal of Heat Transfer-Transactions of the Asme, 2011. **133**(6).
24. Dames, C. and G. Chen, *Theoretical phonon thermal conductivity of Si/Ge superlattice nanowires.* Journal of Applied Physics, 2004. **95**(2): p. 682-693.
25. Debye, P., *The theory of specific warmth.* Annalen Der Physik, 1912. **39**(14): p. 789-839.
26. Chen, G. and T.F. Zeng, *Nonequilibrium phonon and electron transport in heterostructures and superlattices.* Microscale Thermophysical Engineering, 2001. **5**(2): p. 71-88.
27. Duda, J.C., et al., *Extension of the diffuse mismatch model for thermal boundary conductance between isotropic and anisotropic materials.* Applied Physics Letters, 2009. **95**(3).
28. Chen, Z., et al., *Anisotropic Debye model for the thermal boundary conductance.* Physical Review B, 2013. **87**(12).
29. Auld, B.A., *Acoustic fields and waves in solids*. 2nd ed.. ed. 1990, Malabar, Fla.: Malabar, Fla. : R.E. Krieger.
30. Li, H.K., W.D. Zheng, and Y.K. Koh, *Anisotropic model with truncated linear dispersion for lattice and interfacial thermal transport in layered materials.* Physical Review Materials, 2018. **2**(12).
31. Wirtz, L. and A. Rubio, *The phonon dispersion of graphite revisited.* Solid State Communications, 2004. **131**(3-4): p. 141-152.
32. Nemanich, R.J. and S.A. Solin, *1st-Order and 2nd-Order Raman-Scattering from Finite-Size Crystals of Graphite.* Physical Review B, 1979. **20**(2): p. 392-401.
33. Ferrari, A.C., et al., *Raman spectrum of graphene and graphene layers.* Physical Review Letters, 2006. **97**(18).

34. (a) Yan, J., et al., *Electric field effect tuning of electron-phonon coupling in graphene.* Physical Review Letters, 2007. **98**(16); (b) Das, A., et al., *Monitoring dopants by Raman scattering in an electrochemically top-gated graphene transistor.* Nature Nanotechnology, 2008. **3**(4): p. 210-215.
35. Pirkle, A., et al., *The effect of chemical residues on the physical and electrical properties of chemical vapor deposited graphene transferred to SiO2.* Applied Physics Letters, 2011. **99**(12).
36. Geick, R., C.H. Perry, and Rupprech.G, *Normal Modes in Hexagonal Boron Nitride.* Physical Review, 1966. **146**(2): p. 543-&.
37. (a) Nemanich, R.J., S.A. Solin, and R.M. Martin, *Light-Scattering Study of Boron-Nitride Micro-Crystals.* Physical Review B, 1981. **23**(12): p. 6348-6356; (b) Arenal, R., et al., *Raman spectroscopy of single-wall boron nitride nanotubes.* Nano Letters, 2006. **6**(8): p. 1812-1816; (c) Song, L., et al., *Large Scale Growth and Characterization of Atomic Hexagonal Boron Nitride Layers.* Nano Letters, 2010. **10**(8): p. 3209-3215.
38. (a) Chiu, M.H., et al., *Spectroscopic Signatures for Interlayer Coupling in MoS2-WSe2 van der Waals Stacking.* Acs Nano, 2014. **8**(9): p. 9649-9656; (b) Lui, C.H., et al., *Observation of interlayer phonon modes in van der Waals heterostructures.* Physical Review B, 2015. **91**(16); (c) Wang, K., et al., *Interlayer Coupling in Twisted WSe2/WS2 Bilayer Heterostructures Revealed by Optical Spectroscopy.* ACS Nano, 2016. **10**(7): p. 6612-6622; (d) Jin, C., et al., *Interlayer electron-phonon coupling in WSe2/hBN heterostructures.* Nat Phys, 2017. **13**(2): p. 127-131.
39. Smith, G.C., *Surface analysis by electron spectroscopy measurement and interpretation*, in *Updates in applied physics and electrical technology*. 1994, Plenum Press,: New York. p. 1 online resource (xi, 156 pages).
40. Lee, S.M. and D.G. Cahill, *Influence of interface thermal conductance on the apparent thermal conductivity of thin films.* Microscale Thermophysical Engineering, 1997. **1**(1): p. 47-52.
41. (a) Koh, Y.K., et al., *Heat Conduction across Monolayer and Few-Layer Graphenes.* Nano Letters, 2010. **10**(11): p. 4363-4368; (b) Yang, J., et al., *Thermal conductance imaging of graphene contacts.* Journal of Applied Physics, 2014. **116**(2); (c) Foley, B.M., et al., *Modifying Surface Energy of Graphene via Plasma-Based Chemical Functionalization to Tune Thermal and Electrical Transport at Metal Interfaces.* Nano Letters, 2015. **15**(8): p. 4876-4882; (d) Zheng, W.D., et al., *Achieving Huge Thermal Conductance of Metallic Nitride on Graphene Through Enhanced Elastic and Inelastic Phonon Transmission.* Acs Applied Materials & Interfaces, 2018. **10**(41): p. 35487-35494.
42. Koh, Y.K., et al., *Heat-Transport Mechanisms in Superlattices.* Advanced Functional Materials, 2009. **19**(4): p. 610-615.
43. Chen, Z., et al., *Thermal contact resistance between graphene and silicon dioxide.* Applied Physics Letters, 2009. **95**(16).
44. Xinxia, L., et al., *Thermal conduction across a boron nitride and SiO 2 interface.* Journal of Physics D: Applied Physics, 2017. **50**(10): p. 104002.
45. Aljishi, R. and G. Dresselhaus, *Lattice-Dynamical Model for Graphite.* Physical Review B, 1982. **26**(8): p. 4514-4522.
46. Bosak, A., et al., *Elasticity of hexagonal boron nitride: Inelastic x-ray scattering measurements.* Physical Review B, 2006. **73**(4).
47. (a) Stassis, C., et al., *Lattice-Dynamics of Hcp-Ti.* Physical Review B, 1979. **19**(1): p. 181-188; (b) Serrano, J., et al., *Vibrational properties of hexagonal boron nitride: Inelastic X-ray scattering and ab initio calculations.* Physical Review Letters, 2007. **98**(9).

48. Chen, G., *Thermal conductivity and ballistic-phonon transport in the cross-plane direction of superlattices.* Physical Review B, 1998. **57**(23): p. 14958-14973.
49. Wolfe, J.P., *Imaging phonons : acoustic wave propagation in solids*. 1998, Cambridge, U.K. ; New York: Cambridge University Press. xiii, 411 p.
50. Huang, B. and Y.K. Koh, *Negligible Electronic Contribution to Heat Transfer across Intrinsic Metal/Graphene Interfaces.* Advanced Materials Interfaces, 2017. **4**(20).
51. Chen, G., *Nanoscale energy transport and conversion : a parallel treatment of electrons, molecules, phonons, and photons*. MIT-Pappalardo series in mechanical engineering. 2005, New York: Oxford University Press. xxiii, 531 p.
52. Tohei, T., et al., *Debye temperature and stiffness of carbon and boron nitride polymorphs from first principles calculations.* Physical Review B, 2006. **73**(6).

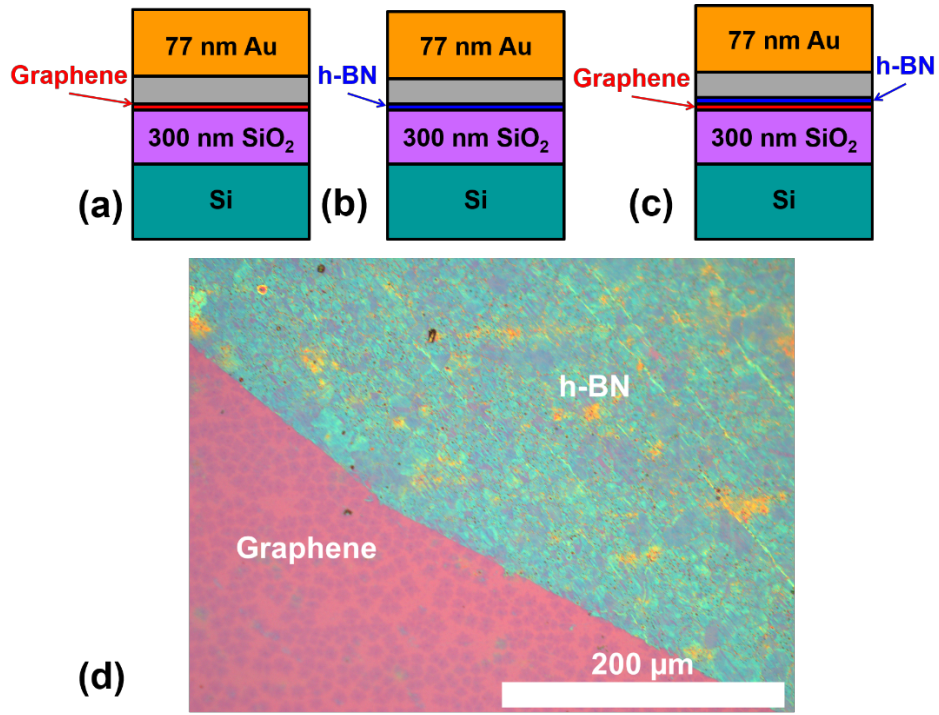

Figure 1: The samples used in this study are CVD grown (a) graphene, (b) h-BN, and (c) h-BN/G. Samples were coated with a Au thermal transducer (3 nm Ti adhesion layer) for TDTR measurements. The interfaces are considered Ti/G/SiO$_2$ or Ti/h-BN/SiO$_2$ in accordance with Schmidt et al. [15] where a 5 nm Ti adhesion layer nearly doubled the TBC at Al-graphite interface. (d) An optical microscope image (20x) showing a ~0.5x0.5 mm$^2$ area on the surface of h-BN/G sample.

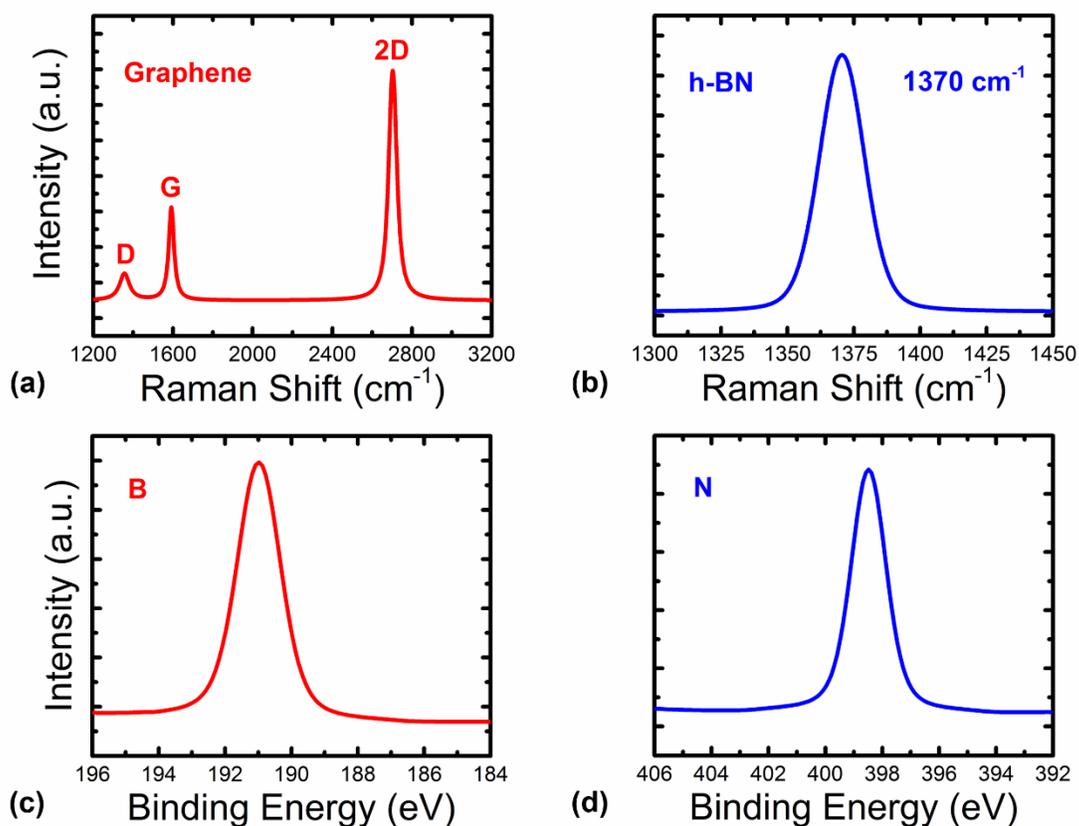

Figure 2: (a) Graphene and (b) h-BN Raman spectra. The intensity ratio I(2D)/I(G) ≈ 2.2 [33] in (a) indicates graphene sample is single-layer. High resolution XPS spectra for h-BN samples showing (c) B and (d) N peaks at 191 and 398 eV, respectively. From the XPS data, we determined the stoichiometry of our h-BN sample was 1.17:1 (B:N).

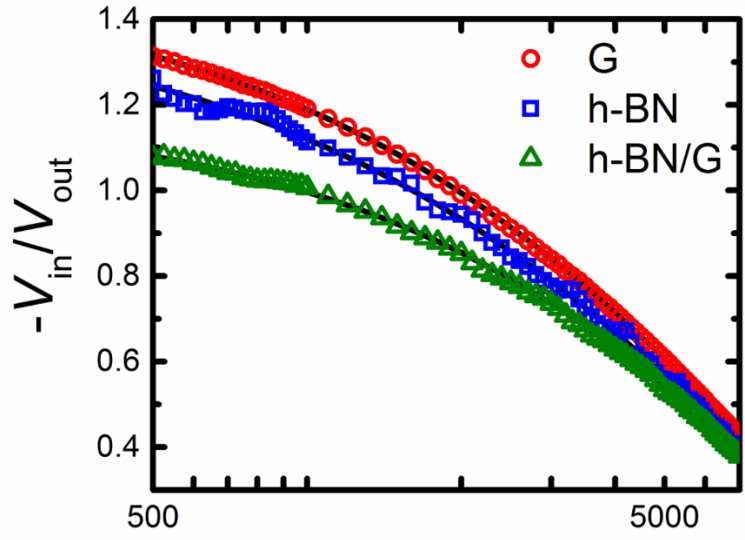

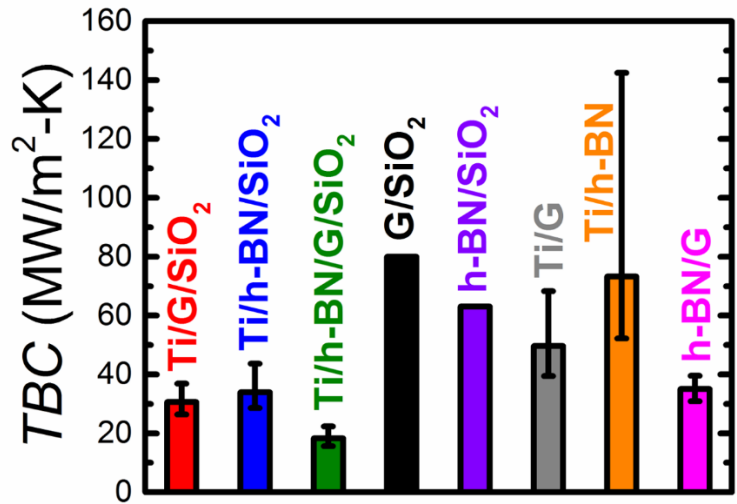

Figure 3: (a) TDTR signal comparisons for three samples used in this study. (b) Summary of TBC results from TDTR measurements and thermal resistor network. Ti/G/SiO$_2$, Ti/h-BN/SiO$_2$, and Ti/h-BN/G/SiO$_2$ values measured using TDTR. Error bars were calculated using a Monte Carlo method [18a]. G/SiO$_2$ and h-BN/SiO$_2$ data taken from references [43] and [44], respectively. Ti/G, Ti/h-BN, and h-BN/G TBC values estimated using series resistance approximation.

Table 1: Input parameters for A-DMM and PWA-DMM models. The phonon velocities were calculated using the elastic constants for Ti, graphite, and h-BN and cutoff frequencies for each branch were determined from the published dispersion relations [31, 47]. The wavevectors, $k_{ab,1}$ and $k_{ab,2}$, frequencies $\omega_{ab,1}$ and $\omega_{ab,2}$, and phonon velocities $v_{ab,1}$ and $v_{ab,2}$, for TL2 branch were determined using the analysis of Li et al. [30]. $v_{ab,2}$, $\omega_{ab,1}$, and $k_{ab,1}$ are not used in the A-DMM model.

| Branch | Parameters | Graphene | h-BN |
|---|---|---|---|
| TA | $v_c$ (m/s) | 1329 | 1915 |
| | $v_{ab}$ (m/s) | 13935 | 12364 |
| | $\omega_c$ ($10^{12}$ rad/s) | 8.24 | 9.98 |
| | $\omega_{ab}$ ($10^{12}$ rad/s) | 190 | 166 |
| TL1 | $v_c$ (m/s) | 1329 | 1915 |
| | $v_{ab}$ (m/s) | 21628 | 19652 |
| | $\omega_c$ ($10^{12}$ rad/s) | 8.24 | 9.98 |
| | $\omega_{ab}$ ($10^{12}$ rad/s) | 231 | 202 |
| TL2 | $v_c$ (m/s) | 4013 | 3586 |
| | $v_{ab,1}$ (m/s) | 1329 | 1915 |
| | $v_{ab,2}$ (m/s) | 7485 | 4396 |
| | $\omega_c$ ($10^{12}$ rad/s) | 22.8 | 22.8 |
| | $\omega_{ab,1}$ ($10^{12}$ rad/s) | 5.34 | 4.08 |
| | $\omega_{ab,2}$ ($10^{12}$ rad/s) | 100 | 61.7 |
| | $k_{ab,1}$ ($10^{10}$ m$^{-1}$) | 0.402 | 0.213 |
| | $k_{ab,2}$ ($10^{10}$ m$^{-1}$) | 1.83 | 1.80 |

Table 2: Fitted phonon transmission coefficients, $\alpha_{12,\text{fit}}$, used in DMM analysis determined by fitting to room-temperature TDTR data.

| Interface | A-DMM [27] | PWA-DMM [28] |
|---|---|---|
| Ti/G | 0.05424 | 0.05424 |
| Ti/h-BN | 0.08016 | 0.08016 |
| h-BN/G | 0.02494 | 0.2287 |

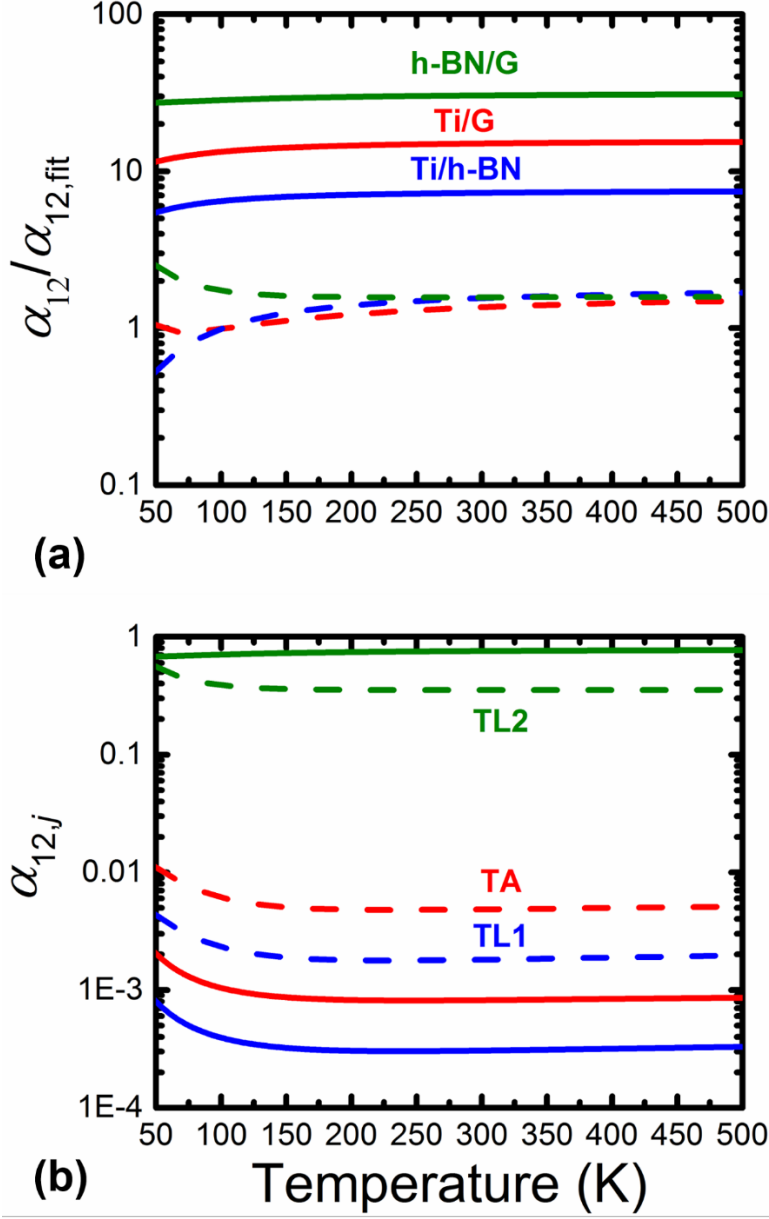

Figure 4: (a) Ratio of transmission coefficients, $\alpha_{12}/\alpha_{12,\text{fit}}$, for A-DMM (solid lines) and PWA-DMM (dashed lines), where $\alpha_{12}$ is calculated from phonon irradiation (Equations 2 and 3) and the relationship $\alpha_{12} = H_2/(H_1 + H_2)$ and $\alpha_{12,\text{fit}}$ is determined from RT TDTR data. The ratio $\alpha_{12}/\alpha_{12,\text{fit}}$ depend weakly on temperature above 200 K. (b) The transmission coefficients, $\alpha_{12,j}$, of different phonon branches (TA, TL1, TL2) as a function of temperature highlights the importance of the TL2 branch to the total transmission $\alpha_{12}$ for h-BN/G interface. Here, solid lines are for A-DMM and dashed lines for PWA-DMM.

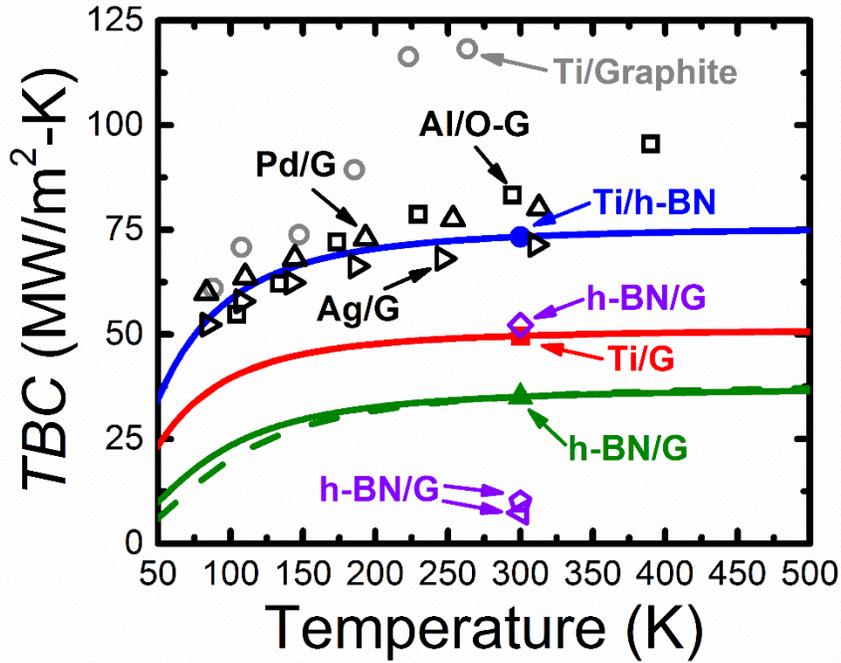

Figure 5: Comparison of TBC for Ti/G (filled red square), Ti/h-BN (filled blue circle), and h-BN/G (filled green triangle) interfaces from this work. DMM results from this work are plotted as solid (A-DMM) and dashed lines (PWA-DMM). TBC values at Ti/G, Ti/h-BN, and h-BN/G interfaces are estimated assuming series resistances. The DMM results were calculated using $\alpha_{12,\text{fit}}$ values in Table 2. Also shown are previously reported of h-BN/G TBC from Chen et al. [12b] (open left purple triangle), Liu et al. [12d] (open purple diamond), and Kim et al. [12e] (open purple trapezoid) using Raman spectroscopy. For further comparison, TBC for various metal/G [Al/O-G (open black square) [16f], Pd/G (open up black triangle) [50], and Ag/G (open right black triangle)[41d]] and Ti/graphite (open gray circles) [15] interfaces are also shown.